\documentclass{article}
\usepackage{graphicx} 
\usepackage{myart}
\usepackage{authblk}
\newcommand{\toolname}{EVSEFuzz }

\usepackage{fancyhdr}

\title{\vspace{-1cm}\toolname: A Tool for Automated Testing of Protocol Implementations}

\date{}

\begin{document}
\thispagestyle{plain}

\author[1]{Soham Gumaste}
\affil[1]{Ohio State University\footnote{Contributions made while affiliated to University of Illinois Chicago}}
\affil[]{\textit{gumaste.1@osu.edu}}
\author[2]{Michael Khan}
\affil[2]{University of Illinois Chicago}
\affil[]{\textit{mkhan398@uic.edu}}
\author[3]{Roland Varriale}
\affil[3]{Argonne National Laboratory}
\affil[]{\textit{rvarriale@anl.gov}}
\author[4]{Rigel Gjomemo}
\affil[4]{University of Illinois Chicago}
\affil[]{\textit{rgjome1@uic.edu}}

\maketitle

\begin{abstract}
The rapid adoption and development of electric vehicle (EV) networks has required the expansive deployment of EV charging infrastructure (EVCI) which relies on communication protocols such as the open charge point protocol (OCPP) to coordinate data transfer between charging network operators and the EV supply equipment. In this paper, we present EVSEFuzz, a tool for automatically building fuzzers to test the security of protocol implementations.  To achieve this we make a novel use of Python datatype validation packages and tools to ``understand'' datatypes and message structures and generate both compliant and non-compliant messages. To date, EVSEFuzz has been tested successfully with several open and closed-source implementations and has found several vulnerabilities leading to the publication of two CVEs. 
\end{abstract}


\section{Introduction}

 The recent adoption of EVs has led to a rapid development of EV charging infrastructures (EVCI). Because of the novelty of this phenomenon, the EVCI environment is fragmented into many stakeholders, including charging companies, car manufacturers, power companies, and customers. In turn, such fragmentation makes it challenging to create common security models and techniques that apply across the industry. In this context, while a few widely accepted protocols exist, like Open Charge Point Protocol (OCPP), Open Charge Point Interface (OCPI), their implementations may vary widely across the EV charging industry. Specifically, the rapid expansion of the role of OCPP within modern charging applications has the ability to amplify the consequences of attacks through OCPP due to the increased reach and degree of connectivity. This includes expansion of Vehicle-to-Everything (V2X) applications to include distributed energy resources, eMobility systems, and home energy distribution.  In this fragmented environment, it becomes important to be able to create security testing methods that can scale across multiple implementations with little manual efforts.
 
 Several works have focused on identifying potential challenges and pitfalls to OCPP implementations \cite{garofalaki2022electric}; however, more generalized testing and assessment methods have been proposed by only one paper \cite{boussaha2025emuocpp}; however, this framework does not include fuzzing analysis of software implementations of communication protocols. 
 
 In this paper, we present EVSEFuzz, a tool that is used to build fuzzers for any remote communication protocol. Given a specification of the messages in the protocol as a JSON schema or an XML schema, EVSEFuzz is able to automatically build compliant and non compliant messages that can be sent to a testing target remotely. To address potential concerns related to intellectual property, \toolname is implemented as a black-box fuzzing approach that does not require access to the code. As part of a larger project sponsored by Argonne National Laboratory and the Department of Energy, we have evaluated \toolname on software belonging to one EV charging company (with whom we signed an NDA), and several open source software projects. \toolname found several non-compliance cases and zero-day vulnerabilities resulting in two CVE entries. To date, \toolname is able to fuzz implementations of OCPP (all versions), OCPI 2.2.1, and OpenADR and can be expanded to work with additional communication protocols with open specifications. 

\section{\toolname  Architecture}

Above all else, the development of \toolname serves to answer a huge problem that plagues the security of communications utilizing any protocol: how do you ensure that {\em any} given implementation of said protocol is secure? This includes customized implementations of such protocols, which may deviate from the specifications. Work previously performed by Argonne National Laboratory has identified the lack of secure web socket implementation on EVSE as a security gap. Manual fuzzing is possible, but implementing a different fuzzer for each protocol implementation is cost-intensive. EVSEFuzz automates the fuzzer creation for each protocol implementation starting from its specifications. 

\begin{figure}[!ht]
     \centering
     \includegraphics[width=\linewidth]{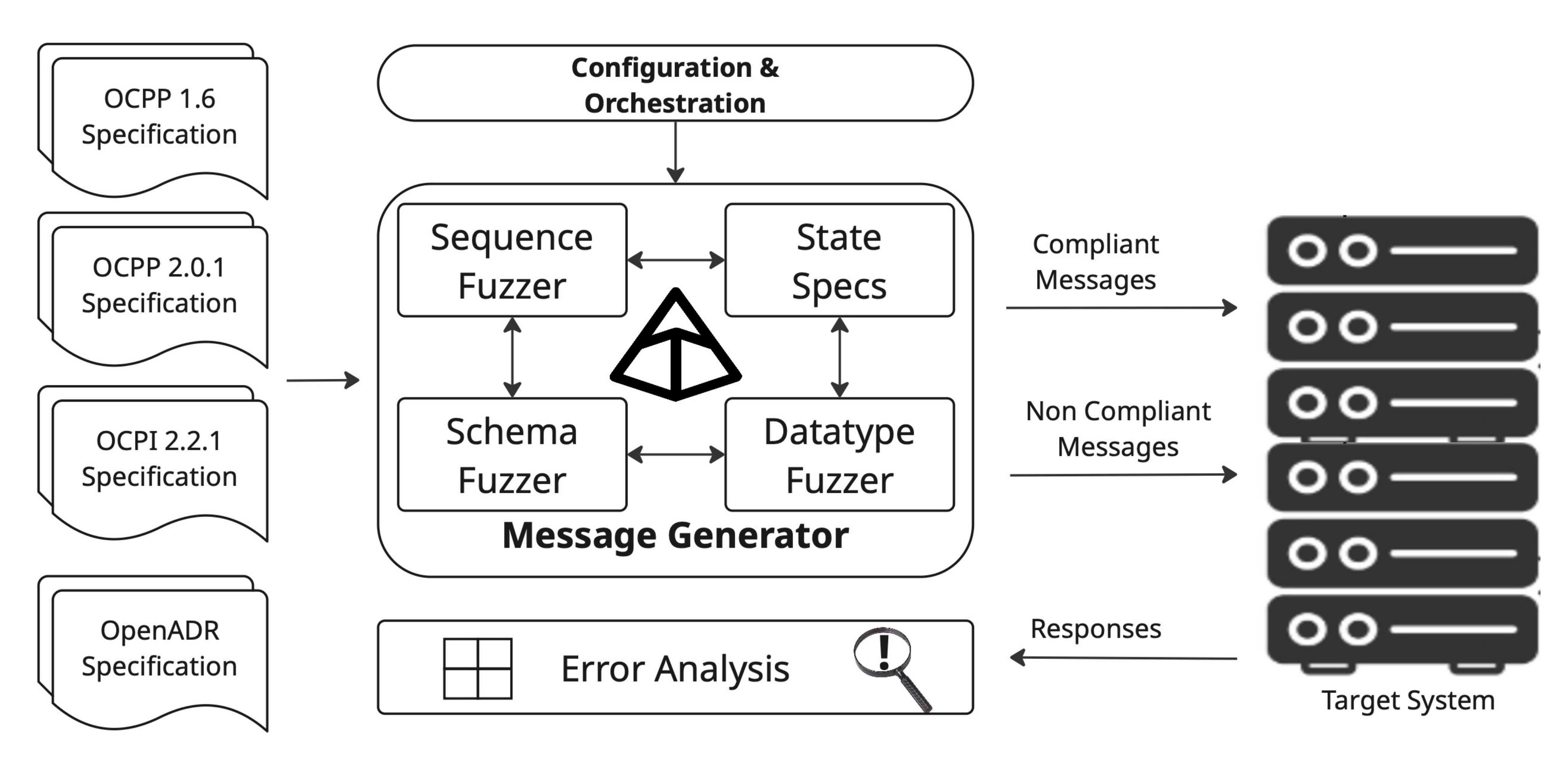}
     \caption{EVSEFuzz Architecture.}
     \label{fig:architecture}
 \end{figure}

A schematic figure of our tool is shown in Figure~\ref{fig:architecture}. The tool receives input from a set of JSON or XML schemas that specify the message format for a specific communication protocol implementation. EVSEFuzz is standard-agnostic - these schemas may conform with a specific standard or deviate from it. EVSEFuzz uses these specifications to guide the creation of both compliant and non compliant messages to send to endpoints implementing that protocol. A configuration and orchestration file describes the endpoint and the sequence of fuzzing operations, which creates different message streams corresponding to different fuzzing strategies. 

\subsection{Datatype Fuzzer} 

This module is able to create valid and invalid messages dynamically from the protocol specification files. This module is implemented with a novel use of the popular runtime data validation library, PydanticT \cite{pydanticRepo}. Messages are intentionally distorted to produce invalid messages in the following ways: 1) Creating values for message fields that are longer than the length specified in the protocol (e.g., a 5000 bytes-long value for a 20 bytes-long field), 2) Changing the data types of values in fields (e.g., alphabetic characters instead of numbers), 3) Supplying values outside the constraints required by the formal protocol definition (e.g., negative numbers where positive numbers are specified in the protocol).

For each of the above, the fuzzer may randomly choose to follow the constraints of the protocol, resulting in a diverse combination of invalid and valid fields in any given message. Similarly, the fuzzer may be tweaked to produce fully compliant messages with randomized values.  

For many of these data types there also exist additional methods which generate data which violate specific rules or test certain edge cases. For example, the Random Fuzzer has methods to generate strings which are longer than allowed by protocol specifications and which contain common escape characters. In these instances the fuzzer could discover compliance violations which could have become critical vulnerabilities.

Pydantic \cite{pydanticRepo} is a popular data validation library for Python. in Python \noindent \textbf{Using Pydantic for Automated Message Generation}.
EVSEFuzz uses Pydantic to reason about datatypes specified in the JSON and XML schema files and create corresponding messages \cite{pydanticRepo}. Pydantic is a popular Python data validation library that helps fill the gap left by the lack of types and type-safety by providing APIs and tooling for modeling and verifying data at runtime. 
Pydantic provides a rich type definition system with support for arbitrary restrictions or constraints. For example, a programmer may specify a composite data type whose field 'x' must be an integer between 0 and 10, or a string field that must match a particular regular expression, and so on. Pydantic integrates well with the type-hints introduced in modern versions of Python. 

We introduce a novel use of this powerful feature-set to generate messages of arbitrary forms. In our implementation, Pydantic models exist as runtime objects in Python, and can thererfore be manipulated at runtime using Python's robust metaprogramming interface. These models are built by importing the protocol specification files. Once imported as Pydantic models, EVSEFuzz recursively walks over the heirarchy of types and constraints, populating values along the way. For example, if a particular field is a list of a composite type, the fuzzer will choose an arbitrary length for the list (including a length within the constraints) and then recurse into the composite type to produce those many values. In a given composite type, primitive values are filled according to their type and constraints, for example a 'Date' field is populated with a appropriate or fuzzed Date specification, or an 'integer' field is filled with a random number and so on. Using Pydantic in this way enables EVSEFuzz to scale to virtually any protocol specification that can be provided as JSON and XML schemas. 

\subsection{Schema and Sequence Fuzzers}

These modules stress test a protocol implementation's  ability to enforce messages' adherence to the structure provided by protocol specifications. Specifically, the {\em schema fuzzer} creates non-compliant messages that violate the message structure specification, e.g., failure to include message fields required by the specification, addition of unsupported message fields, repetition of fields with different values, etc. These messages will always at minimum have one field and at most have seven. These limits were decided to emulate the typical length of a message sent by a client. Each of these fields' value is then set to a random value for either a primitive data type or an object, whose fields are similarly primitive data types set to a random value.

The {\em sequence fuzzer} is responsible for scheduling the sequence of messages that are exchanged with a remote endpoint. In particular, this module is able to send a compliant sequence of messages (e.g., to perform authentication) followed by non-compliant messages or messages that are not expected (e.g., instead of sending a Response message to a client in response to a client's request, the fuzzer may send a Request message).

\noindent
\textbf{State Specification}. One of the major drawbacks of using an arbitrary fuzzing strategy is that coverage of the protocol implementations may be limited. In fact, just by sending randomly generated messages, it is very unlikely to reach deeper states in the communication protocol. To deal with this limitation, the different modules use also {\em State Specifications}, which are simple text files containing a sequence of messages that can be sent to a target endpoint to be able to reach a specific state in the communication session. For instance, authentication messages are included in this file, to authenticate the session and be able to test post-authentication message implementations. When configured to do so, EVSEFuzz reads these messages from the file and sends them. 

In addition to authentication messages, this state specification can also contain messages with more targeted content, including malicious payloads (e.g., cross-site scripting, SQL injection) for the message fields. These payloads test the robustness of the overall software architecture of the application against untrusted input sources. For instance, it is common for backend server operations of charging companies to have web interfaces that retrieve data from backend databases that are populated with inputs from the charging stations. In this mode, we include malicious SQL injection payloads in our messages to test the database implementation robustness, and malicious cross-site scripting payloads to test the robustness of the web-based operator interfaces. Payloads for other types of attacks are also possible (e.g., command injection, memory-based attacks, etc). This allows EVSEFuzz to be able to test against a wide range of vulnerabilities and preemptively discover vulnerabilities, which a malicious actor may use against a system.

\subsection{Error Analysis and Configuration}
\noindent 
\textbf{Error Analysis}. Because it is built as a black-box tool, the only feedback that \toolname receives from the remote protocol implementation are the responses to the messages sent to the test target. 
To discover anomalous behaviors, EVSEFuzz includes an {\em Error Analysis} module. Specifically, for each message sent, we analyze the response and check whether that response is compliant with the expected response in the protocol. For instance, if a non-compliant message does not trigger an error response message, it means that the protocol implementation may have accepted that non-compliant message. On the other hand, if a compliant message triggers an error response, it means that the target is encountering errors with benign input and may be an indication of a bug. \toolname flags these messages and their timestamps to share with the customers who have access to the code and can determine the root cause. In tests over open-source implementations of OCPP, we are able to quickly pinpoint the root cause of the error using the type of message and error information returned by the application.

\noindent
\textbf{Configuration and Orchestration}. This component of the tool deals with fuzzing strategies that are specific to each target. Specifically, for each target, we specify the endpoint IP address and port, testing times and frequency, based on customers' specifications, protocols to be tested, and files containing protocol specifications, state specifications, certificates, etc. More importantly, this module drives the tool to execute different fuzzing strategies in different phases. The most common sequence of phases includes the following: 1) Authentication messages using the sequence fuzzer, 2) A possibly empty sequence of compliant messages to establish a connection and reach a specific state (e.g., charging session started) using the sequence fuzzer, and 3) Random or state-based fuzzing, where non-compliant messages or messages with malicious payloads are sent to the target. In particular, the random fuzzing involves invoking both the Schema and Datatype fuzzers to generate messages with strings of random characters. In the Schema fuzzer, a random number generator then determines how many additional fields are added to the message.

    In both of these cases, the junk data that each field is set to is not always entirely random. In many cases, the random fuzzer may choose to instead populate a field with a specific type of data known to reveal non-compliance in many systems. For example, if a string field is chosen, the random fuzzer is capable of filling the field with special control characters. If a target system is not properly sanitize their inputs, this could lead to a system crash where otherwise random junk data would have been properly rejected.

\textbf{Error Analysis}. For every execution of \toolname, a record of the fuzzing attempt is made in the results folder provided by the user. All the tests executed for a given day are stored in a file named \texttt{stats-time-MM-DD-YY}, wherein the last three hyphenated values are two digit representations of the month, day, and year of execution respectively. Additional attempts are appended to this file and properly denoted for ease of analysis later.
Additionally, tThe starting and end times of each execution of the fuzzer, alongside the sending and reciept of messages during the fuzzer's connection with the target system. All communications are fully logged.

There are two main ways errors arise that can be documented in \toolname's logs. System responses and the connection between the fuzzer and target system being severed. At the moment, checking the validity of system responses to fuzzer messages requires manual analysis, but can be done quickly due to the validity of the sent message being denoted before its matching response. Analyzing these logs is an efficient way to find when a target system marked a valid as invalid and an invalid message as valid. On the other hand, when the connection between the fuzzer and its target is severed, the last message which was sent before the connection error is logged and denoted as the potential cause for the connection being closed. It should be noted that this does not mean that the message was responsible for this interruption. However, in the case that the fuzzer was the cause, it suggests that there is non-compliance within a system that may allow malicious attackers to disable or exploit it.

\section{Evaluation and Results}

 sure,  to it, as you shouldthatThe idea is that twould make ir,but that theyso

\begin{table}[ht!]
\scriptsize
\centering
\begin{tabular}{|l|l|l|}
\hline
\textbf{Error Type} & \textbf{Description} & \textbf{CVE ID} \\
\hline
Out of bounds error in libocpp.~\cite{libocpp1} & A larger than expected string in a JSON   & CVE-2025-59399~\cite{cve1} \\
                                                & field causes a crash in EVerest leading   &                            \\
                                                & to potential denial of service attack.     &                            \\
\hline
Malformed input causes crash in                 & Malformed input causes a library used by   & CVE-2025-59398~\cite{cve2} \\
\texttt{everest-framework}.~\cite{everest1}     & EVerest to crash, leading to potential    &                            \\
                                                  & denial of service attack.               &                            \\
\hline
Out of bounds crash in libocpp.                  & A larger than expected string in a JSON   & CVE-2025-59398~\cite{cve2} \\
Fix of first bug introduces this              & field causes a crash in EVerest leading    &                            \\
bug.~\cite{everest2}                          & to potential denial of service attack.     &                            \\
\hline
Uncaught exception causes crash.~\cite{everest3}   & An uncaught exception occurs for larger   & CVE-2025-59398~\cite{cve2} \\
                                                & than expected input strings. The exception      &                            \\
                                                & causes EVerest to crash, leading to potential     &                       \\
                                                & denial of service attack.                         &                       \\
\hline

\end{tabular}
\caption{\small Errors Found Using EVSEFuzz.}
\label{tab:errors}
\end{table}

We have evaluated \toolname on a range of OCPP implementations including EVerest~\cite{everest} (a Linux Foundation Energy project), CitrineOS~\cite{citrineOS}, and the server-side OCPP 1.6 implementation of an EV charging company, with whom we signed an NDA, in the context of a larger project with Argonne National Laboratory. Thanks to its automated approach, in all cases, deployment was fairly straightforward amounting to just pointing the fuzzer at an IP address and setting up the fuzzing strategies orchestration configuration. Specifically, we connected \toolname \space to our partner's endpoint and set the fuzzer to run and send 10,000 random messages two to three times a week. 

While we cannot comment on the results found in our partner EV charging company, we found several vulnerabilities in EVerest, as shown in Table \ref{tab:errors}. Because EVerest is open source, we were able to also pinpoint the root cause by correlating the errors with the source code location. These results are reported in Table \ref{tab:errors}. The most common error that we found is related to out of bounds access and subsequent crashing of EVerest. Additionally, we found several examples of non-compliance, particularly when the datatype fuzzer generated edge case messages. For example, the presence of the value \texttt{NaN} in fields that expected an integer consistently caused the system to falsely accept invalid messages as valid. In many cases, the the schema fuzzer would omit one or more required fields for a message, causing the server to not reject the message, but crash and subsequently sever the fuzzer's connnection to it. It was discovered that during the validation process of a message, the server was assuming a non-existent data field did, in fact exist. By attempting to access this non-existent data, the system crashed.

\section{Conclusion}
In this paper, we describe EVSEFuzz, a practical fuzzing tool for testing the security implementations of communication protocols in the EV charging infrastructure. Given a protocol's specification, EVSEFuzz is able to automatically instantiate a fuzzer for that protocol, cutting development time and simplifying deployment in the field. EVSEFuzz has been tested successfully with the infrastructure of an EV charging company and open-source software. recent line of ourthusly
\bibliographystyle{IEEEtran}
\bibliography{references}
\end{document}